\documentclass[conference]{IEEEtran}
\IEEEoverridecommandlockouts
\usepackage{cite}
\usepackage{amsmath,amssymb,amsfonts}
\usepackage{algorithmic}
\usepackage{graphicx}
\usepackage{textcomp}
\usepackage{xcolor}
\usepackage{multirow} 
\usepackage{xcolor}
\usepackage{hyperref}
\usepackage{subfigure}   

\usepackage[pscoord]{eso-pic}

\newcommand{\placetextbox}[3]{%
  \setbox0=\hbox{#3}%
  \AddToShipoutPictureFG*{%
    \put(\LenToUnit{#1\paperwidth},\LenToUnit{#2\paperheight}){%
      \vtop to 0pt{\null\makebox[0pt][c]{#3}\vss}%
    }%
  }%
}

\def\BibTeX{{\rm B\kern-.05em{\sc i\kern-.025em b}\kern-.08em
    T\kern-.1667em\lower.7ex\hbox{E}\kern-.125emX}}
\begin{document}

\title{Spectrum Configuration Framework for Throughput Maximization in Open Systems with Roll-Off-Based QoT Optimization\\
}



\author{
\IEEEauthorblockN{Peyman Pahlevanzadeh\IEEEauthorrefmark{1}\IEEEauthorrefmark{3},
Venkata Virajit Garbhapu\IEEEauthorrefmark{1},
Agastya Raj\IEEEauthorrefmark{1},
Dmitrii Briantcev\IEEEauthorrefmark{2}, \\
Dan Kilper\IEEEauthorrefmark{2},
Marco Ruffini\IEEEauthorrefmark{1}}
\IEEEauthorblockA{\IEEEauthorrefmark{1}School of Computer Science and Statistics, ADAPT centre, Trinity College Dublin, Ireland}
\IEEEauthorblockA{\IEEEauthorrefmark{2}School of Engineering, CONNECT Centre, Trinity College Dublin, Ireland}
\IEEEauthorblockA{\IEEEauthorrefmark{3}\textit{pahlevap@tcd.ie}}
}
\maketitle

\placetextbox{0.2}{0.055}{978-3-903176-78-2 \textcopyright\ 2026 IFIP}

\begin{abstract}
We propose a spectrum-configuration framework for open and disaggregated optical
systems that maximizes throughput while guaranteeing the quality of transmission
(QoT) margins. The framework jointly optimizes transceiver parameters, including
modulation format, symbol rate, pulse-shaping roll-off factor, and wavelength-
selective switch (WSS) bandwidth, under fixed spectral allocation constraints.
The impact of roll-off factor optimization is first experimentally evaluated in
the presence of cascaded WSS filtering, demonstrating measurable QoT gains for
both single- and multi-channel transmission. Building on these observations, a
knapsack-based optimization is applied in the context of Optical Spectrum as a
Service (OSaaS) to select service configurations that maximize aggregate
throughput within a fixed spectrum width and limited transceiver resources.
Experimental validation on a metro-scale open testbed confirms the effectiveness
of the proposed approach in achieving efficient spectrum utilization and
adaptive throughput–margin trade-offs.
\end{abstract}



\section{Introduction}

Optical networks are evolving from fixed-grid architectures toward elastic
optical networks (EONs), driven by the increasing demand for data-intensive
applications and the need for more efficient spectrum utilization. Unlike fixed
50 or 100~GHz channel spacing, EONs based on flex-grid allow variable-width
channels, offering more flexibility. Flexible wavelength-selective switches
(flex-WSSs) enable dynamic bandwidth allocation at the channel level
\cite{7522660}. Building on this flexibility and the increased interest in open
and disaggregated systems, operators are exploring Optical Spectrum as a Service
(OSaaS), in which tenants lease spectrum slices and independently manage their
transceivers, with the potential of supporting optical interconnection across
heterogeneous domains \cite{10750159}.

A large number of prior work has focused on optimizing the line system, including
the configuration of in-line Erbium-Doped Fiber Amplifiers(EDFAs), gain equalization \cite{10024064}, and launch
power optimization along the optical link \cite{8896064}. These approaches primarily target
link-level impairments and aim to maximize performance margins by tuning
amplifier settings and transmitted power. In contrast, the impact of
transceiver-level configuration parameters has received comparatively less
attention in open and disaggregated optical systems.

A key challenge in this context is achieving high spectral efficiency without
compromising quality of transmission (QoT). The roll-off factor (ROF) of the
root-raised-cosine (RRC) pulse-shaping filter remains underexplored despite its
critical impact. Increasing the ROF reduces inter-symbol interference (ISI) but
also expands the occupied bandwidth \cite{Harako:16, Pahlevanzadeh:24}. In
addition, the ROF directly affects tolerance to residual impairments such as
skew \cite{Yue:18}. This trade-off is particularly critical for higher-order
modulation formats, such as DP-64QAM, where increased ISI sensitivity can
severely degrade QoT. Selecting appropriate ROF and WSS bandwidth can therefore
mitigate filtering penalties and improve performance margins without requiring
additional hardware.
Proprietary systems typically rely on vendor-specific planning and engineering
tools to determine optimized system and transceiver configurations. In contrast,
open tools for disaggregated systems, such as GNPy \cite{9768871}, primarily focus
on transmission performance and generally lack mechanisms for optimizing
transceiver parameters such as modulation format, symbol rate, and pulse-shaping
roll-off factor. In this work, we first experimentally evaluate
the impact of ROF on QoT for different modulation formats and symbol rates under
WSS filtering constraints. We then introduce a spectrum-configuration framework
for OSaaS that jointly optimizes transceiver parameters to configure a fixed
spectral allocation with guaranteed QoT margins while maximizing overall
throughput. The proposed approach provides a practical, measurement-driven
solution for efficient and high-throughput operation in OSaaS-based EONs.

\section{Methodology}

\begin{figure}
\centering
\includegraphics[width=0.80\columnwidth]{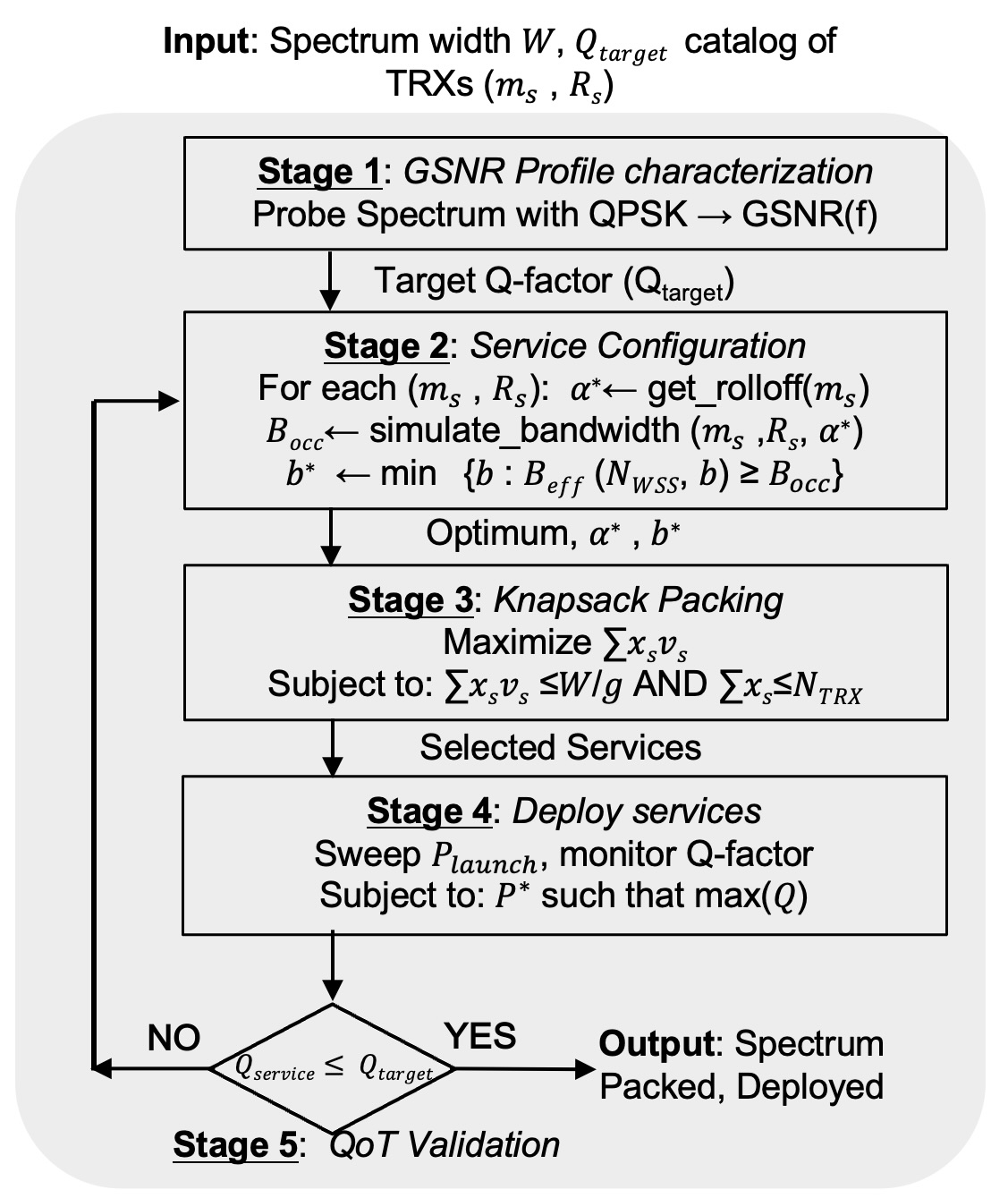}
\caption{Workflow of the proposed spectrum-configuration framework for OSaaS and open line systems, comprising GSNR profiling, service configuration, knapsack-based spectrum packing, deployment with launch-power optimization, and QoT monitoring.}

\label{fig:flow_CHART}
\end{figure}

In OSaaS or open line systems (OLS), a tenant acquires a fixed spectral band (e.g., $W = 300$ GHz) and either the tenant (OSaaS) or operator (OLS) must configure it to maximize throughput while guaranteeing QoT. The tenant has: (i) a service catalog of modulation formats $m_S$ and symbol rates $R_s$, (ii) $N_{\text{TRx}}$ transceivers, (iii) topology information (number of Reconfigurable Optical Add/Drop Multiplexers(ROADMs) and EDFAs, and, where possible, (iv) a Generalized Signal-to-Noise Ratio (GSNR) spectral profile from QPSK probing. All selected channels must meet $Q_{\text{target}}(m) = Q_{\text{FEC}}(m) + \Delta Q_{\text{margin}}$\cite{7830257}. Our framework comprises five stages (Fig.~\ref{fig:flow_CHART}).

\textbf{Stage 1: GSNR Profiling.} We sweep a QPSK probe signal across the allocated
spectrum \cite{10122551}, measure the frequency-dependent BER, and convert it to a GSNR profile, $\mathrm{GSNR}(f)$. Given the target margin
$\Delta Q_{\text{margin}}$, this GSNR profile is used to determine the set of
feasible services that can be supported at each spectral location. Specifically,
for each modulation format and symbol rate in the service catalog, we verify
whether the corresponding QoT requirement $Q_{\text{target}}$ can be satisfied.
This process identifies, across the spectrum under test, which combinations of modulation format and symbol rate are eligible for deployment under the desired
QoT margin.

\textbf{Stage 2: Service Configuration.} For each feasible service $(m_s, R_s)$, we determine two components: (i) optimal ROF $\alpha^*(m)$ from pre-characterized transceiver data (vendor-provided or experimentally obtained), and (ii) \emph{minimum WSS bandwidth} $b^*$ via simulation using the WSS model~\cite{Pulikkaseril:11}. The simulator estimates the occupied bandwidth $B_{\text{occ}}(R_s, \alpha^*)$ through time-domain signal generation and power spectral density computation, then uses a cascade model $B^{(3\text{dB})}_{\text{eff}}(N_{\text{WSS}}, b)$ to select the smallest $b^*$ satisfying $B_{\text{3dB}}(N_{\text{WSS}}, b^*) \geq B_{\text{occ}}(R_s, \alpha^*)$. This ensures the signal fits within the effective passband after $N_{\text{WSS}}$ cascaded filters without clipping. Each service is recorded as $(m_s, R_s, \alpha^*, b^*)$ with throughput 
$v_s = 2R_s\log_2(M(m_s))$ and bin cost $k_s = \lceil b^*/g\rceil$, 
where $g = 6.25$ GHz is the WSS granularity. 
The throughput $v_s$ represents the gross symbol rate; the net client data rate accounts for 27\% SD-FEC overhead. 
Since the SD-FEC overhead is fixed at 27\% across all services, 
we omit this factor from the optimization formulation without loss of generality.

\textbf{Stage 3: Knapsack Packing.} The available spectrum is discretized into
$B = \lceil W/g \rceil$ frequency bins, and the service selection problem is
formulated as a two-dimensional bounded knapsack optimization. The objective is
to maximize the aggregate throughput $\sum x_s v_s$ subject to a spectrum
constraint $\sum x_s k_s \le B$ and a transceiver constraint
$\sum x_s \le N_{\text{TRx}}$. The problem is solved using dynamic programming,
with the state DP$[b][t]$ representing the maximum achievable throughput using
$b$ spectrum bins and $t$ transceivers.

\begin{figure}
    \centering
    \includegraphics[width=0.90\linewidth]{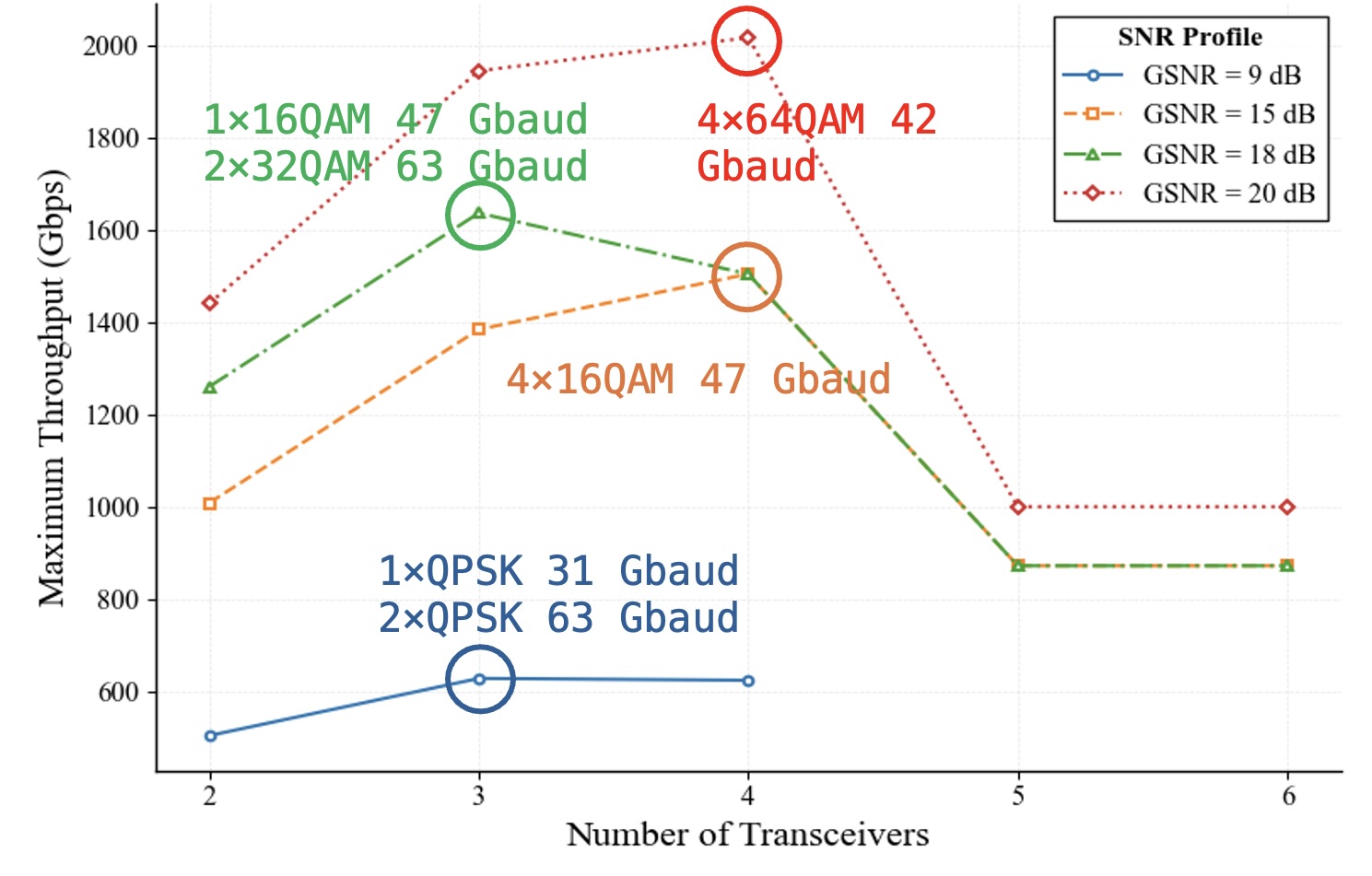}
   
    \caption{Maximum throughput versus number of transceivers for different GSNR profiles of the link in the fixed 200 GHz spectrum. Circles indicate the service configurations (modulation format and symbol rate) selected by the knapsack optimization.}

    \label{fig:num_trx_knapsack}
\end{figure}

Fig.~\ref{fig:num_trx_knapsack} illustrates the output of the knapsack
optimization, showing the maximum achievable throughput versus the number of transceivers for different GSNR profiles within a fixed 200~GHz spectral window.
The circles indicate the service configurations selected by the optimizer,
including the modulation format and symbol rate determined in Stage~2. An
important observation is that increasing the number of transceivers does not
necessarily lead to higher throughput. For example, for a GSNR profile of
18~dB, deploying three transceivers yields a higher total throughput than using four transceivers, as the additional channel must operate at a more conservative
service configuration to satisfy the QoT constraint. This highlights the
benefit of jointly optimizing spectrum allocation and transceiver parameters,
rather than maximizing the number of deployed channels.

\textbf{Stage 4: Deployment \& Power.} After deploying the selected channels, we sweep the transmitter booster gain and measure the Q-factor of all channels. In OSaaS systems, users have no control over in-line EDFAs along the lightpath; thus, launch power optimization is limited to the booster EDFA at the transmitter side.

\textbf{Stage 5: Monitoring.} After deploying the selected channels, we monitor the Q-factor of all channels. 
If $Q_{\text{measured}} < Q_{\text{target}}$, the spectrum allocation is deemed infeasible, and Stage~2 is re-run to 
select the next-best feasible service configuration. This iterative process continues until the QoT requirement is satisfied. 

    
    

\subsection*{Optimization Problem Formulation}

Let $\mathcal{S}$ denote the set of feasible services obtained from Stage~2. 
Each service $s \in \mathcal{S}$ is characterized by a throughput
$v_s = 2 R_s \log_2(M_s)$ and a required WSS bandwidth $b_s$.
The corresponding spectrum cost in discrete bins is
$k_s = \left\lceil b_s / g \right\rceil$, where $g$ denotes the WSS granularity.
The total available spectrum is $W$, yielding
$B = \left\lceil W / g \right\rceil$ spectrum bins, and
$N_{\text{TRx}}$ transceivers are available.

We define the binary decision variable
\begin{equation}
x_s =
\begin{cases}
1, & \text{if service $s$ is selected}, \\
0, & \text{otherwise}.
\end{cases}
\end{equation}

The optimization problem is formulated as
\begin{align}
\max_{\{x_s\}} \quad & \sum_{s \in \mathcal{S}} v_s x_s
\label{eq:obj} \\
\text{s.t.} \quad
& \sum_{s \in \mathcal{S}} k_s x_s \le B,
\label{eq:spectrum_constraint} \\
& \sum_{s \in \mathcal{S}} x_s \le N_{\text{TRx}},
\label{eq:trx_constraint} \\
& x_s \in \{0,1\}, \quad \forall s \in \mathcal{S}.
\label{eq:binary_constraint}
\end{align}

Only services satisfying the QoT requirement
$Q_{\text{measured}} \ge Q_{\text{target}}$ are included in $\mathcal{S}$.
Consequently, QoT constraints are enforced implicitly by construction.

\begin{figure}
    \centering
    \includegraphics[width=0.76\linewidth]{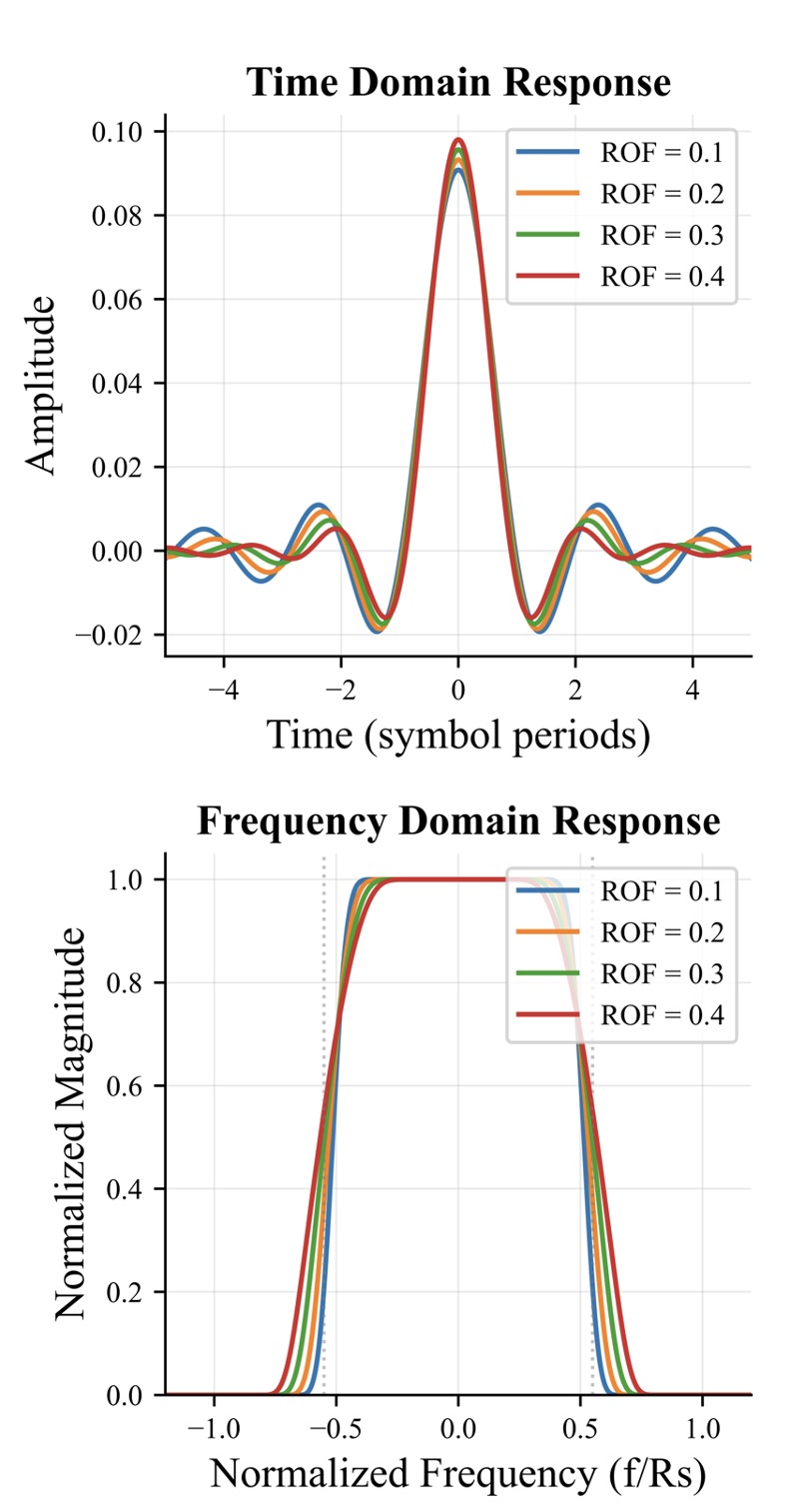}
    \caption{RRC pulse shaping in the time and frequency domains for different roll-off factors.}

    \label{fig:ROF_time_frequncy}
\end{figure}

 \begin{figure*}[t]
    \centering

    \subfigure[ Topology A]{
        \includegraphics[scale=0.3]{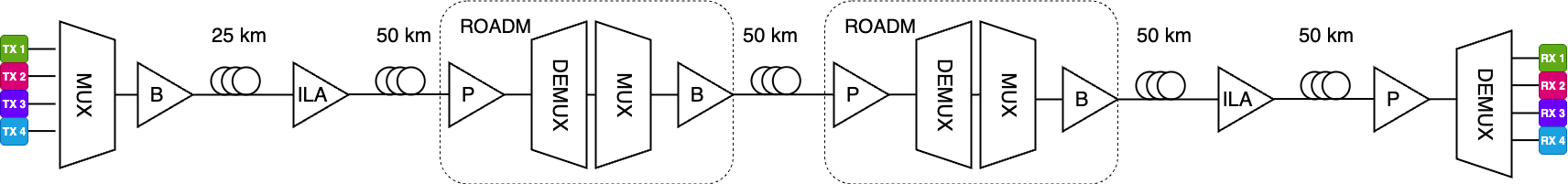}
        \label{fig:topo_B}
    }

    \vspace{2mm}

    \subfigure[ Topology B]{
        \includegraphics[scale=0.3]{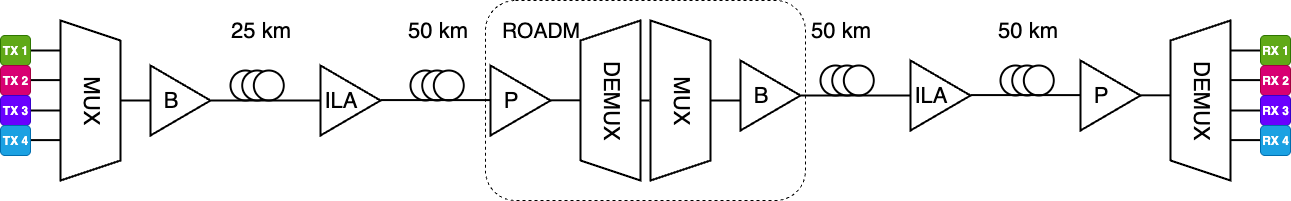}
        \label{fig:topo_A}
    }

   \caption{Topologies used to evaluate ROF optimization and spectrum-configuration performance.
    (a) Five-span configuration with 6 WSSs.
    (b) Four-span configuration with 4 WSSs.}
    
    \label{fig:topo_B}
\end{figure*}

\begin{figure}
    \centering
    \includegraphics[scale=0.85]{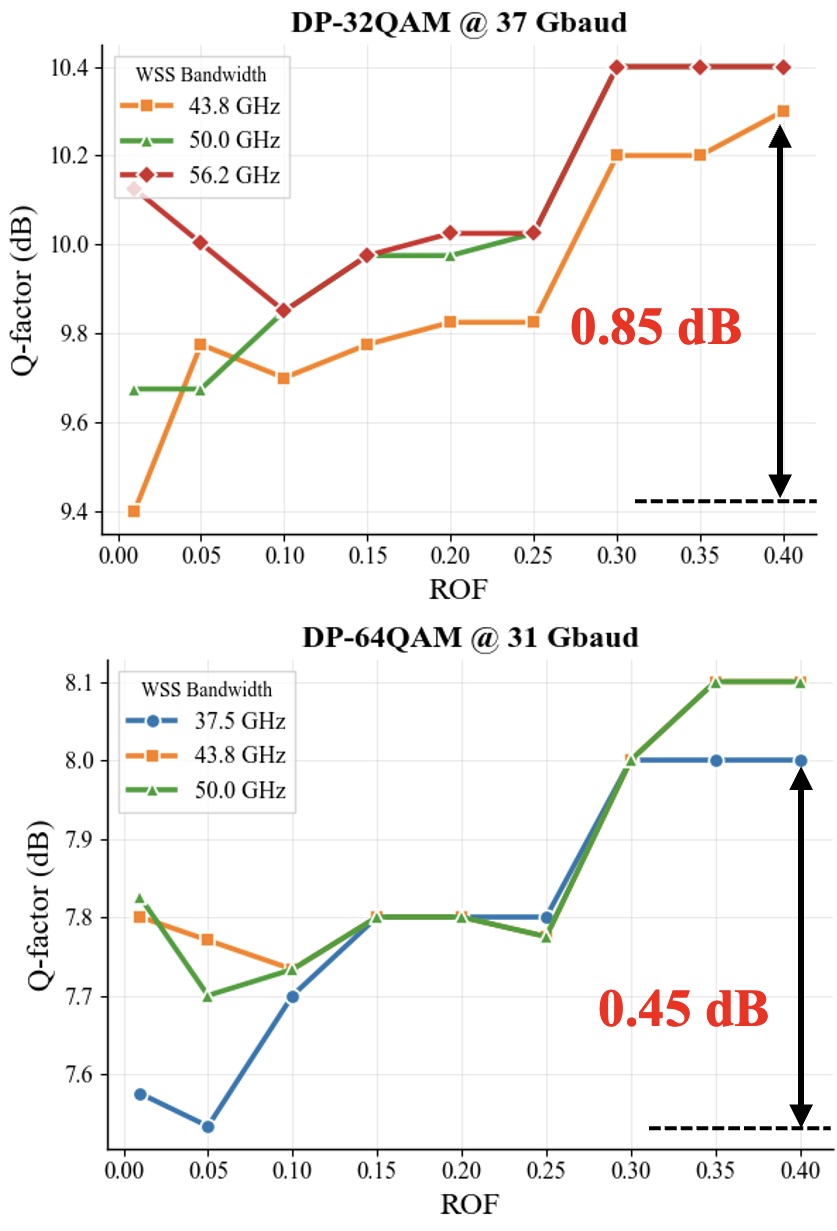}
    \caption{Measured Q vs. ROF for single channel DP-64QAM@31 GBd and DP-32QAM@37 GBd across WSS bandwidths}
    \label{fig:single_channel}
\end{figure}

\section{Experimental Setup}
\label{sec:experimental}
To characterize the impact of ROF optimization in the presence
of WSS filtering, we first consider a single-channel experimental setup. The
testbed comprises a commercial Adtran Teraflex coherent transceiver employing
RRC pulse shaping at the transmitter, followed by a MUX-side WSS, a booster EDFA,
70~km of standard single-mode fiber, a pre-amplifier EDFA, a DEMUX-side WSS, and a
receiver with standard DSP. This configuration allows controlled evaluation of
how the transmitted signal, constrained by a finite WSS passband, is affected by
changes in ROF The experimental characterization of single-channel ROF optimization and its impact on QoT is presented and discussed in
Section~\ref{sec:results}.

In the second step as shown in Fig.~\ref{fig:topo_B}(b), we extend the analysis to a multi-channel scenario to evaluate
the effectiveness of ROF optimization in an OSaaS context. We deploy four
Teraflex transceivers within a fixed 300~GHz spectral window and consider a
metro-style topology comprising a MUX and DEMUX at the transmitter and receiver,
respectively, and one intermediate ROADM to introduce additional filtering
effects. The total link length is approximately 175~km. In this scenario, we
jointly optimize the ROF and the launch power of the transmitter-side booster EDFA
for all four channels. Finally, to evaluate the framework under more severe filtering conditions as shown in Fig.~\ref{fig:topo_B}(a), we
consider a third scenario with two ROADMs along the lightpath, resulting in a
total of six cascaded WSSs. This configuration further accentuates bandwidth-narrowing effects and is used to validate the proposed workflow under strong filtering constraints. The experimental results for this scenario are also
presented and discussed in Section~\ref{sec:results}.

\begin{figure*}[t]
    \centering
    \subfigure[DP-16QAM optimum configuration]{
        \includegraphics[width=0.48\linewidth]{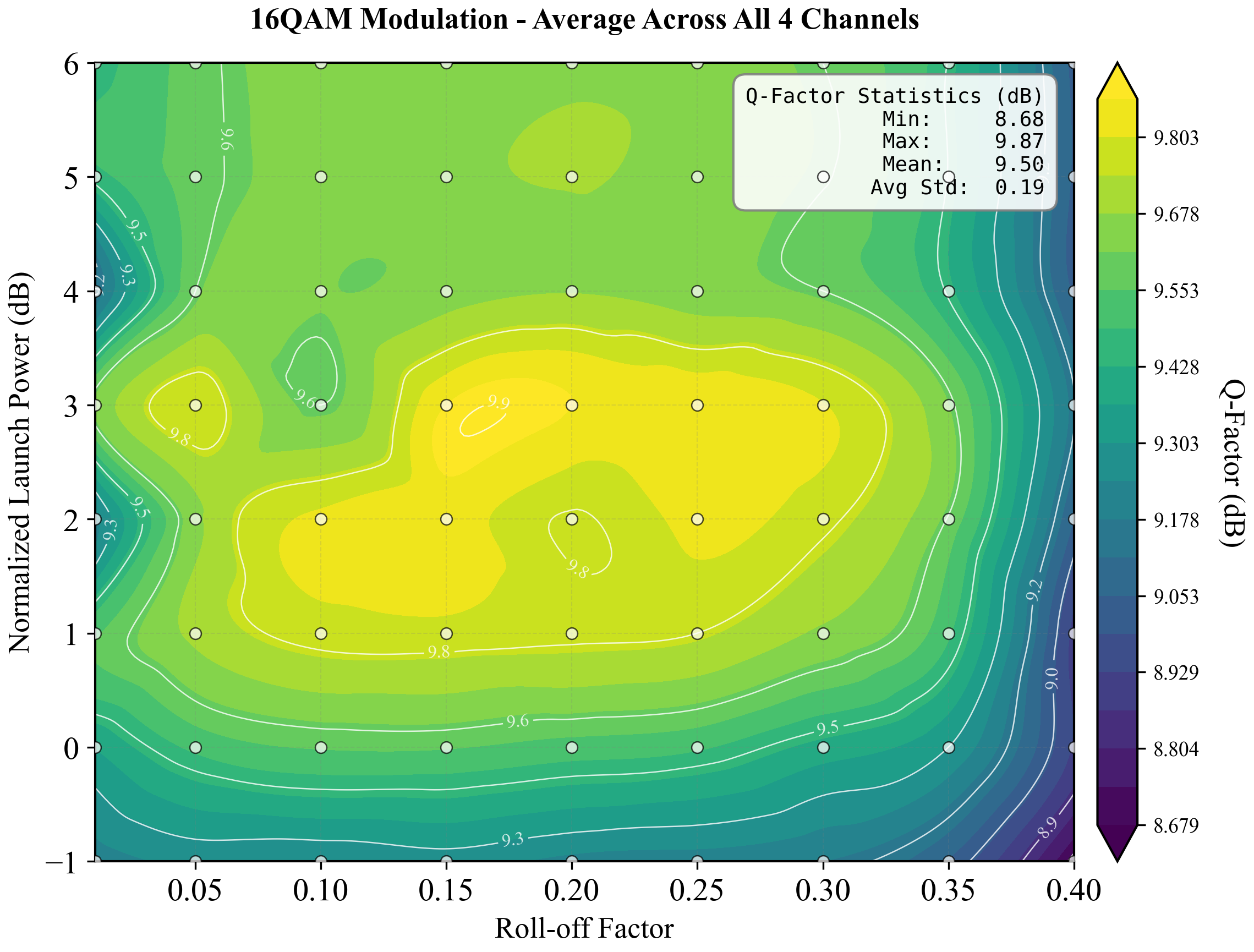}
        \label{fig:16qam_opt}
    }
    \hfill
    \subfigure[DP-32QAM optimum configuration]{
        \includegraphics[width=0.48\linewidth]{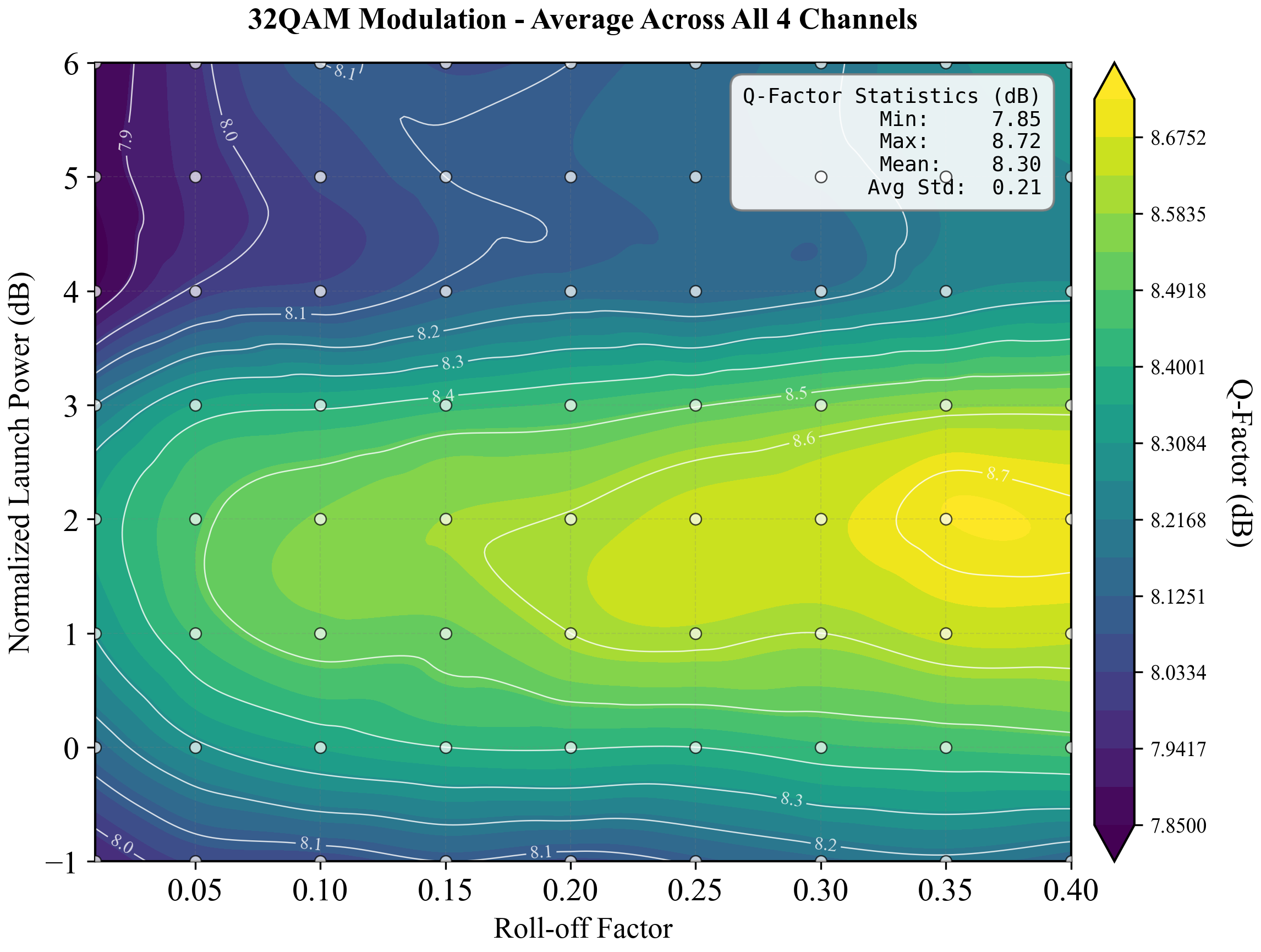}
        \label{fig:32qam_opt}
    }
    \caption{Optimum roll-off and launch-power operating points under WSS filtering for (a) DP-16QAM and (b) DP-32QAM.}
    \label{fig:contour_plots}
\end{figure*}


\begin{table*}[t]
\centering
\caption{Experimental results of the spectrum under test(SUT) configurations.}
\scriptsize
\setlength{\tabcolsep}{3.5pt}
\renewcommand{\arraystretch}{1.1}
\begin{tabular}{cccccccccc}
\hline
\textbf{Knapsack Opt.} & \textbf{SUT (GHz)} & \textbf{\#TRXs} & \textbf{GSNR (dB)} & 
\textbf{Mod.} & \textbf{Sym. Rate (Gbaud)} & \textbf{ROF} & 
\textbf{WSS BW (GHz)} & \textbf{Q-Fact (dB)} & \textbf{Cap. (Gbps)} \\
\hline
\multirow{4}{*}{4\texttimes64QAM--42G} 
 & 200 & 4 & 19.91 & \textbf{64QAM} & \textbf{42} & 0.35 & 50 & 6.87 & \textbf{2016} \\
 & 200 & 4 & 19.88 & 64QAM & 31 & 0.35 & 50 & 6.87 & 1488 \\
 & 200 & 4 & 18.97 & 32QAM & 37 & 0.30 & 50 & 9.05 & 1480 \\
\hline
\multirow{4}{*}{4\texttimes32QAM--63G} 
 & 300 & 4 & 17.85 & \textbf{32QAM} & \textbf{63} & 0.20 & 75 & 7.90 & \textbf{2520} \\
 & 300 & 4 & 19.53 & 64QAM & 52 & 0.40 & 75 & 6.40 & 2496 \\
  & 300 & 4 & 16.08 & 16QAM & 63 & 0.15 & 75 & 9.17 & 2016 \\
\hline
\end{tabular}
\vspace{-3pt}
\label{Table_SUT}
\end{table*}

\section{Results and Discussion}
\label{sec:results}
Fig.~\ref{fig:ROF_time_frequncy} illustrates the effect of the ROF on RRC pulse shaping in the time and frequency domains. The top plots show the impulse responses of the RRC filter for different ROF values. As the ROF increases from 0.1 to 0.4, the temporal sidelobes of the pulse decay more rapidly, indicating reduced ISI. The bottom plots of Fig.~\ref{fig:ROF_time_frequncy} present the corresponding frequency responses of the RRC filter. Increasing the ROF leads to a wider occupied bandwidth, reflecting the trade-off between time-domain compactness and spectral efficiency. In the presence of cascaded WSS filtering, this bandwidth expansion can increase susceptibility to filtering penalties. Therefore, understanding the joint impact of ROF on both time-domain compactness and spectral occupation is essential for evaluating system performance under realistic WSS-constrained transmission conditions. Fig.~\ref{fig:single_channel} presents the measured Q-factor as a function of the
ROF for different WSS bandwidths, obtained from the single-channel experiments conducted using the setup described in
Section~\ref{sec:experimental}. The results are shown for DP-32QAM at 37~GBd
(top) and DP-64QAM at 31~GBd (bottom). Increasing the ROF leads to improved Q-factor for both modulation formats. In particular, for DP-32QAM, ROF optimization yields
up to 0.85~dB improvement in Q-factor under narrow WSS filtering, while for
DP-64QAM a maximum gain of approximately 0.45~dB is observed. 

 The contour plots shown in Fig.~\ref{fig:contour_plots}(a) present the average Q-factor of four channels tested in the topology of Fig.~\ref{fig:topo_B}(b) with four cascaded WSSs. We assumed a 300~GHz spectral budget and loaded four channels. In \emph{Scenario~1}, each channel carried DP-16QAM at 63~Gbaud, mapped to a 75~GHz WSS window with zero guard band. In \emph{Scenario~2}, each channel carried DP-32QAM at 50~Gbaud, again using a 75~GHz WSS window across all MUX/DEMUX elements along the lightpath. The optimization jointly tuned the ROF and the transmitter booster launch power. The contour plots illustrate the joint dependence of Q-factor on relative launch power and ROF, where relative launch power denotes the difference between the actual launch power and the optimum launch power. Fig.~\ref{fig:contour_plots}(a) shows the measured Q-factor trends for DP-16QAM due to its lower-order modulation, ISI sensitivity is reduced, and the best operating point occurs at a moderate roll-off of \(\mathrm{ROF}\approx [0.15,0.2]\). For DP-32QAM shown in Fig.~\ref{fig:contour_plots}(b), ISI is more pronounced, and increasing the ROF improves QoT; the optimum occurs at \(\mathrm{ROF}\approx [0.35,0.4]\). In both scenarios, the Q-factor improves with increasing launch power until NLI becomes dominant, beyond which further increases degrade QoT. Table~\ref{Table_SUT} presents the experimental results obtained from the 6-WSS topology shown in Fig.~\ref{fig:topo_B}(a). The first row for each SUT represents the optimal configuration predicted by the knapsack optimizer. Additional modulation format and symbol rate combinations were tested around these configurations to verify that the solutions were indeed optimal. As observed, the configurations achieving the highest throughput align with the optimizer's predictions. For the 200~GHz allocation, the optimizer correctly identifies DP-64QAM at 42~Gbaud as the maximum-capacity solution (2016~Gbps). However, with a measured Q-factor of 6.87~dB, this configuration operates near the FEC threshold. Suppose the user's margin requirement is not satisfied. In that case, the framework iteratively selects the next-best solution: DP-32QAM at 37~Gbaud, which provides a Q-factor of 9.05~dB with a sufficient margin at a throughput of 1480~Gbps, demonstrating the adaptive throughput-margin trade-off.

\section{Conclusions}
We presented a spectrum-configuration framework for OSaaS and open line systems that maximizes throughput by jointly optimizing transceiver parameters, including modulation format, symbol rate, and pulse-shaping roll-off factor. Experimental validation on the OpenIreland testbed demonstrates that proper ROF tuning yields measurable Q-factor improvements under cascaded WSS filtering, with gains up to 0.85~dB. The knapsack-based optimization effectively balances throughput and margin requirements, enabling adaptive service selection within fixed spectral allocations. The proposed framework provides a practical, measurement-driven approach for tenants and operators to maximize spectral efficiency in optical networks without additional hardware investment.



{\footnotesize
\textbf{Acknowledgments. } Work supported by research grants from Research Ireland under 22/FFP-A/10598, 18/RI/5721, 13/RC/2077 P2, 13/RC/2106 P2.
}

\bibliographystyle{IEEEtran}
\bibliography{samples}

\end{document}